%% file: paper.tex
\documentclass[sigconf, nonacm]{acmart}

\usepackage{graphicx}
\usepackage{tabularx}

\def\BibTeX{{\rm B\kern-.05em{\sc i\kern-.025em b}\kern-.08emT\kern-.1667em\lower.7ex\hbox{E}\kern-.125emX}}
    
\copyrightyear{2024}
\acmYear{2024}
\setcopyright{acmlicensed}
\acmConference[ICSE 2024]{46th International Conference on Software Engineering}{April 2024}{Lisbon, Portugal}
\acmPrice{15.00}
\acmDOI{10.1145/1122445.1122456}
\acmISBN{978-1-4503-9999-9/18/06}

\input{utils}

\begin{document}

%
\title[Objectives and Key Results in Software Teams]{Objectives and Key Results in Software Teams: \\ Challenges, Opportunities and Impact on Development}

%

\author{Jenna Butler}
\affiliation{%
  \institution{Microsoft}
  \country{USA}
}
\email{jennbu@microsoft.com}

\author{Thomas Zimmermann}
\affiliation{%
  \institution{Microsoft}
  \country{USA}  
}
\email{tzimmer@microsoft.com}

\author{Christian Bird}
\affiliation{%
  \institution{Microsoft}
  \country{USA}
}
\email{cbird@microsoft.com}

%
\renewcommand{\shortauthors}{Butler et al.}

%
\begin{abstract}
Building software, like building almost anything, requires people to understand a common goal and work together towards it. In large software companies, a VP or Director will have an idea or goal and it is often the job of middle management to distill that lofty, general idea into manageable, finite units of work. How do organizations do this hard work of setting and measuring progress towards goals? To understand this question, we undertook a mixed methods approach to studying goal setting, management dissemination of goals, goal tracking and ultimately software delivery at a large multi-national software company. \par 
Semi-structured interviews with 47 participants were analyzed and used to develop a survey which was deployed to a multi-national team of over 4,000 engineers. The 512 responses were analyzed using thematic analysis, linear regressions and hypothesis testing, and found that tracking, measuring and setting goals is hard work, regardless of tools used. Middle management seems to be a critical component of the translation of lofty goals to actionable work items. In addition, attitudes and beliefs of engineers are critical to the success of any goal setting framework.
Based on this research, we make recommendations on how to improve the goal setting and OKR process in software organizations: invest in the data pipeline, increase transparency, improve communication, promote learning communities, and a structured roll out of OKRs.
\end{abstract}

%
%


%

%

%
\maketitle


\section{Introduction}
In 1968, Edwin Locke put forward the first academic paper on the theory of goal setting \cite{locke}. Since then, goals and various frameworks for setting and achieving these goals have been used at companies around the world. Many large and successful companies, including Intel, Google, The Gates Foundation, YouTube, Adobe and Intuit specifically use the OKR framework - the Objectives and Key Results framework originally conceived by Andy Grove and popularized by John Doerr in his book Measure What Matters \cite{Doerr01}. The OKR framework is a goal-setting methodology that aligns the objectives of an organization, team, and individual by establishing clear, measurable key results to measure progress towards and achieve strategic outcomes.  It has been successfully used to drive innovation, alignment, and organizational focus, initially at Intel and later at many large companies\cite{Doerr01}. Often, the OKR framework is especially attractive to software companies because OKRs are heavily based in data and measurement - a trait often shared by software companies. While many books exist on how to use OKRs (such as Measure What Matters~\cite{Doerr01}, The OKR Fields Book~\cite{lamorte2022}, OKRs for All~\cite{vellore2022okrs}, Objectives and Key results Leadership~\cite{gray2019}, etc.) there is little research on the use and implementation of such frameworks in software organizations. 

OKRs are said to bring with them 4 superpowers: 1) Focus and commit to priorities 2) Align and connect for teamwork 3) Track for accountability and 4) Stretch for amazing \cite{Doerr01}. Previous work showed 70\% of U.S. employees are disengaged at work, even though progress toward a meaningful goal is a top motivator for employees \cite{HBR2016}. OKRs are purported to help people see the meaningful goal they are working towards, and help people commit to it. 

\added{This paper shares a case study and the experience of a team of over 4,000 engineers that adopted the OKR framework. We used a mixed methods approach and analyzed the data from 47 interviews and 512 survey responses using thematic analysis, linear regression, and hypothesis testing (Section~\ref{sec:methodology}). The following research questions are addressed:

\begin{itemize}
    \item \textbf{RQ1}: What \textbf{behaviors, team practices and work cultures} are associated with a good OKR practice? (Section~\ref{sec:goodOKRculture})
    \item \textbf{RQ2}: What \textbf{challenges} are faced when implementing an OKR system in a large organization? (Section~\ref{sec:challenges})
    \item \textbf{RQ3}: What \textbf{best practices} can improve an OKR process? (Section~\ref{sec:rec})
\end{itemize}
}

\added{This paper contributes a case study that highlights the challenges that SE teams face when adopting the OKR framework. This paper also shares experiences on how a large software team has addressed these challenges and what benefits were realized when adopting the OKR framework. Finally, the paper provides recommendations based on empirical findings for how to improve goal setting in software engineering organizations.}


\section{Background and Related Work}
\subsection{The OKR Framework}
In \textit{The Practice of Management}, Peter Drucker outlined his management principal of ``management by objectives and self control'' \cite{mbo}. This was in strong context to the leading management style of the early 20th century defined by Fredrick Winslow Taylor and Henry Ford - an authoritarian style that did not take into account individuals abilities and treated individuals more like machines \cite{taylor}, \cite{ford}. Management by objectives was a more human way to manage built on respect and trust of the individual.

Andy Grove was working at Intel in the 1960s when Drucker's management style, called ``MBO'' for Management By Objectives, was being used. A meta-analysis study had shown that a high commitment to MBOs did lead to productivity gains of 56\% \cite{rodgers} however there were still imitations to this style - goals were often set at a high level and did not involve those doing the work; they would be set and forgotten, stagnating without updating; and they sometimes were reduced to just numbers, which aren't always very motivating. Grove observed this system in use and came up with his own version titled iMBOs for Intel Management by Objectives. In this new iteration, objectives almost always were coupled with key results; objectives included a what and a how; they moved at a faster pace, being re-evaluated every quarter; were shared publicly; and were often aspirational in nature \cite{Doerr01}. 

In 1975 when John Doerr came to Intel, this version of iMBO was used throughout the company. Doerr nicknamed the system \textbf{OKR} for \textbf{Objectives and Key Results} and came to learn their full power when he was a manager at Intel. Since then, Doerr has popularized the methodology, sharing it with companies such as Google, The Gates Foundation and Intuit \cite{Doerr01}.

The OKR framework can be summarized as follows:
\begin{itemize}
    \item{Every quarter, set 1-3 objectives}
    \item{Objectives are defined as a simple, one sentence goal that begins with a verb and is quantitative. It often takes the form of: ``verb + what you want to do + in order to/for/so what''. An example might be ``Drive better attendance at our conference in order to improve the customer's networking experience''}
    \item{For each objective, identify 2-4 key results. Key results are quantitative values that will provably show if the objective has been met. Key results take the form: ``verb + what you're going to measure + from x to y''. An example would be: ``Increase attendance from 300 to 500 people''.}
    \item{Every month, review each objective and corresponding key results. Highlight any that are at risk and make a plan to change course as needed.}
    \item{At the end of every quarter, reflect on objectives and key results. Score each key result with whether it was met or missed and learn why that happened. Based on the success or failure of key results, decide if the objective has been met. If it has not, keep it until the next quarter, with either the same or new KRs depending on if the current KRs are completed. If the objective has been met, create a new objective for next quarter.}
\end{itemize}

\subsection{Research on the OKR Framework}
While OKRs are most popular in the technology field, research into the use of Objectives and Key Results has occurred across industries, in the fields of education \cite{mangipudi}, \cite{fonseca}, \cite{ruixia}, \cite{rao}, utilities \cite{charo}, \cite{trink}, economics \cite{businessSchool}, and hospitality \cite{pureheart}.

Mangipudi \textit{et al.} believed OKRs could help higher education institutes to enhance communication, engagement and accountability, and presented a framework to help institutions utilize OKRs to get these benefits \cite{rao}. Other studies in the field of education found similar results. Cao looked at doing teaching evaluations using OKRs as well as some big data techniques and found that this method could improve both the quality of education and teaching as well as teachers' initiative \cite{ruixia}. Da Fonseca had a small sample size but believed that using OKRs could be helpful to improve teachers' leaderships kills and give more structure to the teacher/student dialogue about performance \cite{fonseca}. Additionally, Mangipudy \textit{et al.}\ looked at using OKRs at the institution level, not the individual teacher level, to help higher education institutes improve engagement, relationships and accountability \cite{mangipudi}. 

In other fields, such as utilities, Charoenlarpkul and Tantasanee looked at what components of the OKR framework had the largest impact on work performance \cite{charo}. They found that while all elements of the OKR framework are useful and should be considered, the element of ``organization direction'' had the most effect on overall work performance. Trinkenreich \textit{et al.} looked at OKRs in the field of IT \cite{trink}. It can be difficult to tie IT investments directly to business goals, so Trinkenreich \textit{et al.} combined OKRs with another goal strategy, GQM+, to help monitor IT business impact. They found that the OKR framework helped state the goals clearly, while GQM+ helped define the OKRs and tasks needed to achieve them. OKRs have also been compared to another framework, the Key Performance Indicator (KPI) framework in \cite{businessSchool}. They looked at the similarities and differences in these two frameworks, as well as their advantages and disadvantages. Some companies today actually use both frameworks together. Lastly, in the hospitality industry, Irikefe found that using OKRs can have a positive effect on organizational performance of hotels \cite{pureheart}.

\subsection{OKRs in Software Engineering}
In recent years there have been a few studies looking at how OKRs work together with software engineers, especially with agile methodology \cite{ferrazzi}, \cite{stray1}, \cite{stray2}. Stray \textit{et al.} looked at how agile teams make OKRs work through interviews, document review and a survey. They found that OKRs did aide in knowledge sharing and improved transparency between teams (an expected outcome of the OKR framework). They also came up with four strategies that be used to help maximize the OKR framework in the context of large-scale agile teams. 

Stray \textit{et al.} also examined how OKRs work together with Slack in engineering teams. They used a Relational Coordination Theory lens to see how the two tools - one for communication and one for goal setting - affected coordination in distributed large-scale agile. They found OKRs helped the team focus on overall outcomes (this is another purported benefit of the OKR framework) and were used to prioritize work on the team. 

Ferrazzi studied 5 digital start up companies to look for a link between OKRs and the agile methodology \cite{ferrazzi}. They found the use of OKRs brought with it greater flexibility of individual action, great alignment, increase in individual empowerment and actually helped individuals have a more agile mentality. Ferrazzi found that agile methodology is a strong complement to the OKR methodology and suggests the use of OKRs could actually help organizations gain more agility.

\section{Research Setting}
Building software often involves people that are globably distributed, existing legacy software, and teams that came from different places (original teams, acquired teams, newly created teams, etc) and coordination between many autonomous teams has been found to be challenging \cite{Paasivaara}.  In this large software organization (approximately 4,000 engineers and program managers spread across the world), the Vice President asked the first author to analysis why decisions made at the top level of the organization often took a long time to yield changes in the code and why some cross-group projects were successful and quick, while others face resistance and eventually fail. This lead the first author and two research colleagues to do an exploratory study on challenges faced by large software engineering teams in this company. They found that one of the top challenges when building distributed software is goal setting and following through on those goals. This ranges from actually choosing what the goals should be, to measuring the goals, to sharing it out, to working with other teams, to motivating other teams to care about your goals, and to making business and product decisions based on the data coming from measuring those goals. Together these challenges look like those that the objective and key result framework purports to address and in fact, many teams inside the organization were using this framework. As such, the authors decided to study the use of objectives and key results in large software engineering teams.

Challenges such as these do not only exist in this company. For example, prior research found that in large scale agile teams goals are often set at management but don't involve teams and team members do not know what the goals are \cite{moe}. They also found teams struggle with setting and communicating goals \cite{moe}. Prior research by Trinkenreich \textit{et al.} looked at how to help a company create OKRs, as there is not much concrete help on creating measurable key results for objectives \cite{trinkenreich}. Indeed, the proper use of objectives and key results is also a widely felt challenge. In fact, Nicole Forsgren (then VP at GitHub, now researcher at Microsoft) stated: ``OKRs — and especially challenges implementing them — are a big issue. It’s one of the main things I’m asked about when I meet with top customers.'' Indeed, an entire software industry has sprung up to help companies use OKRs more seamlessly, with companies such as WorkBoard, Lattice, ClearPoint and WeekDone all vying for this market. In addition, in 2021 Microsoft purchased OKR software company Ally.io for \$76 million dollars \cite{cnbc} and now sells OKR software under the name Viva Goals. 


\paragraph{Research Questions}
As the organization has been using the OKR framework in some capacity for approximately two years, most teams are familiar with it and have been attempting to use it as their primary approach to setting, tracking, and achieving goals. \added{After creating a measure of OKR maturity, we wanted to understand what behaviors, team practices and cultures were associated with a higher maturity, in order to understand where OKRs would work best and how to improve maturity through a variety of avenues.}

\vspace{4pt}
\noindent{}\textbf{RQ1}: What \textbf{behaviors, team practices and work cultures} are associated with a good OKR practice?
\vspace{4pt}

Management had indicated that there has been mixed success at adoption of the OKR framework across the organization and some teams have had difficulty using the process.
Thus, another goal was to identify and characterize the challenges that teams have faced as they have attempted to use the OKR framework.

\vspace{4pt}
\noindent{}\textbf{RQ2}: What \textbf{challenges} are faced when implementing an OKR system in a large organization?
\vspace{4pt}

As some teams have been more successful in adopting the OKR framework, we seeked to understand what factors, practices, and tools they employed in an effort to share this knowledge with other teams adopting the framework.  

\vspace{4pt}
\noindent{}\textbf{RQ3}: What \textbf{best practices} can improve an OKR process?
\vspace{4pt}

\added{With this research,} we intend to learn best practices and recommendations from software teams who have already attempted to use the framework, drawing on both their successes and failures. \added{The findings from RQ1, RQ2, and RQ3 are discussed in Sections~\ref{sec:goodOKRculture}, \ref{sec:challenges}, and ~\ref{sec:rec} and provide a set of lessons learned and recommended practices for the use of the OKR framework in large software organizations.}

\section{Methodology}
\label{sec:methodology}

In this section we will outline our approach for conducting the initial interviews and subsequent survey design, deployment and analysis. \added{The section is structured into three parts: participants, data collection, and data analysis.}

\subsection{Participants}

\paragraph{Interviews} The study began with an exploratory set of interviews of leaders and individual contributors both inside the organization and with partner organizations that worked with it.
A total of 47 individuals from within and around the organization were chosen for interviews. In order to get good representation, a diverse selection of individuals were chosen with consideration given to the following criteria: global location, level at the company, role in organization and gender. The table \ref{table:1} shows a breakdown of interviews by discipline and level or seniority and the number of each interviewed.

\begin{table}\centering
\caption{Breakdown of interview participants based on role, level and number interviewed. }
\label{table:1}
\begin{tabular} { llc }
 \toprule
 \bf Discipline &  \bf Level of & \bf Number \\
 \bf  & \bf Seniority & \bf Interviewed \\
 \midrule
 Software Engineer  & Entry Level & 4  \\
 Software Engineer  & Senior  & 1 \\
 Software Engineer  & Principal and Above & 4  \\
 Software Engineer  & Manager  & 5  \\
 Software Engineer  & Upper Management  &  11 \\
 \midrule
 Product Manager  & Principal and Above &  3  \\
 Product Manager  & Manager   & 3  \\
 Product Management & Upper Management & 11  \\
 \midrule
 Designer & Principal and Above  & 2  \\
 \midrule
 Data Science & Principal and Above   & 1  \\
 \midrule
 Architect  & Principal and Above  &  3  \\
\bottomrule
\end{tabular}
\end{table}

\paragraph{Survey} The subsequent survey was sent to all members, levels and geographies represented in the organization and a total of 512 responses were collected. Of these, 63 were from entry level engineers and program managers (PMs); 111 were from early career engineers and PMs; 180 were from senior level engineers and PMs; and 153 were from engineers or PMs at Principal level or above. 

\subsection{Data Collection}
\paragraph{Interviews}
The 47 interviews were semi-structured and ranged in time from 30-60 minutes. At least two authors were present for all interviews and interviews were recorded for later analysis. We introduced the topic of wanting to understand what got in the way of moving quickly in the organization and what issues people ran into when working. We then let people answer in an open-ended style and were careful to not seed the discussions with issues brought up by others in order to let each interview stand on its own. Once people exhausted answering our open-ended questions, we would dig more deeply into issues brought up by others if we thought the person being interviewed might have a unique take on it. When interviewing people from outside of the organization, we would ask about issues they faced when working with the organization, or what issues they saw within their own working groups. 

\paragraph{Survey}
We sent the survey to approximately 4,000 employees within the organization we were studying. We used a direct email from the authors email account with a Qualtrics link to solicit responses. We left the survey open for \added{15 days} and reminded individuals once to submit their surveys \added{after 11 days}. We had a survey response rate of approximately 13\% and received 512 responses. 
This response rate is in line with other surveys in software engineering research~\cite{punter2003conducting}.

\added{The survey was advertised to take 15-20 minutes to complete and most participants finished within this time. 
After completion of the survey, participants could enter a sweepstakes to win one of four US \$100 gift certificates, as such rewards have been found to improve response rates~\cite{smith2013improving}.}

\added{The survey included questions about the following aspects:
\begin{itemize}
    \item Demographics (discipline, career stage, experience, region)
    \item Employee engagement and satisfaction (19 items), i.e., how employees feel about their work, how they perceive their rewards and recognition, and their level of autonomy. \newline{\small Example items are \emph{``Most days I am excited to come to work to do my job'', ``My work creates value for customers'',} and \emph{``I am rewarded based on the impact of my contributions to business outcomes''.}}
    \item Team culture (9 items), i.e., the values of open communication, teamwork, and alignment with organizational goals within a team. \newline{\small Example items are \emph{``On my team, new ideas are welcomed'', ``On my team, we are focused on a unified mission'',} and \emph{``On my team we are willing to embrace change''.}}
    \item OKR Maturity (6 items), i.e., how effective the team is at activities related to goal setting.
    \item Modern engineering practices (7 items), i.e., how modern the engineering practices of the team are.
    \item Activities related to goal setting
    \item Communication channels related to goal setting
    \item Time spent on decision making and goal setting
    \item Open ended questions related to challenges, benefits, and improvements with goal setting.
\end{itemize}
The full survey instrument including all questions and all items is available as supplemental materials~\cite{supplemental-materials}.%
}

\subsection{Data Analysis}
\paragraph{Interviews}
After the interviews, we \added{reviewed} the transcripts and recordings for repeated themes. Once we had exhausted theme identification, we categorized which themes appeared in which interviews. We recorded all themes that came up unsolicited in at least two interviews. The list of these themes is presented in the results in Section~\ref{sec:goodOKRculture}.

\paragraph{Survey}
For open text questions on the survey, we used thematic analysis and open coding to group responses into themes. For closed-ended questions we used multiple approaches. 
We primarily utilized linear regression to explore the correlations between different factors of the OKR framework and elements of agile software engineering. 
Additionally, we conducted Fisher’s Exact tests and Mann-Whitney Wilcoxon tests to investigate demographic differences in the survey responses.
All results reported in this paper were statistically significant with a p-value < 0.05.


\section{Good OKR Practices}
\label{sec:goodOKRculture}

\subsection{Interview Themes}
The top themes that were seen in interviews of people both within the organization and those that work with the organization where the following:
\begin{itemize}
    \item Decision making 
    \item Moving together as a broad organization and working across teams
    \item Optimization questions (such as local vs global optimization in a large organization) 
    \item Clarity of goals, mission and alignment between teams
    \item Technical debates (such as writing shareable code vs getting it done quickly, using existing libraries vs writing our own, etc) 
\end{itemize}

Four of the five top themes are purported to be addressed by the Objectives and Key Results framework. OKRs are supposed to reduce fatigue in decision making by clearly stipulating the key results needing to be obtained; help teams work together by making objectives and key results transparent and shared throughout the organization; identify team dependencies early by calling out shared objectives so ``my team or your team'' debates don't ensue; and bring clarity to goals and mission by limiting organizational objectives \added{to a few} top priorities. As such, the authors decided to follow up the interviews with an organization wide survey on the use, challenges and success of OKRs. 

\subsection{OKR Maturity and Experience, Work Culture, and Agile Methodology}
The first set of questions on the survey included demographic information such as gender, length of time at the company, length of time in the organization, length of time in the software industry in general, geographic location and career level. After that we asked general questions about elements of team culture that are related to the OKR framework or adopting to the framework. These included asking about having a unified mission, commitment to organizational priorities and willingness to embrace change. 

In this organization, management and HR also asked yearly questions about the health of the organization and leadership wanted to understand how using the OKR framework correlated with the cultural values they were trying to drive. To this end, we included questions from their annual survey, including questions around spending time doing work you enjoy, being happy at the company, the value of creativity, and the level of challenge at work. Results from all of these questions were later correlated with the OKR maturity score (see below).

\paragraph{OKR Maturity Score}
Using the OKR framework laid out in Measure What Matters \cite{Doerr01}, we broke the framework down into 6 specific pieces. We used the term ``goal'' instead of objective and/or key result throughout much of the survey so that individuals who are not aware of the OKR framework specifically (perhaps more junior engineers who aren't yet involved in business planning) could still respond based on their use of the components of the framework. The 6 elements we identified were as follows:

\begin{itemize}
\item Defining goals
\item Measuring goals
\item Communicating goals
\item Reporting progress towards goals
\item Reviewing progress towards goals
\item Adjusting goals or resourcing towards goals.
\end{itemize}

We asked individuals how effective their team is at these 6 steps of the OKR framework using a Likert scale with options Not Effective at All, Slightly Effective, Moderately Effective, Very Effective and Extremely Effective. 
Using this 6 question mechanism, we then gave each response an \textbf{``OKR Maturity''} score. If individuals selected ``Very Effective'' or ``Extremely Effective'' they scored a 1 for that component of the OKR framework. If they selected anything else, they scored a zero. We summed the score of all 6 components to give each response a maturity score between 0 (not at all mature) and 6 (mature in all levels of the OKR framework). 

\medskip
Maturity Score: 
\begin{equation}
    \sum\limits_{\textit{item}} = 
    \begin{cases}
    1, & \emph{if \emph{item} is \emph{Very Effective} or \emph{Extremely Effective}} \\
    0, & \textit{otherwise}
    \end{cases}
\end{equation}

\paragraph{OKR Maturity Correlations}
Once OKR maturity score was calculated, we used linear regression to examine correlations of OKR maturity with the elements of demographics and team culture mentioned above. Using linear regression with \emph{several} factors allows us to we examine the relationship of each factor to the dependent variable, OKR maturity, while controlling for other factors to isolate and understand the unique contribution of each variable in \added{explaining} OKR maturity.

We found OKR maturity was \emph{positively} correlated with years in industry, having a unified mission as a team, team commitment to organizational priorities, high team value of creativity, spending time doing work that is truly enjoyed, and happiness at the company. 
In contrast, we found OKR maturity to be \emph{negatively} correlated with higher level, years at the organization, being at the main office (vs.\ satellite offices) and the statement ``creating my software is challenging''. 
All of these correlations were statistically significant with a p-value < 0.05. 

\subsection{Modern Development}
In addition to looking at OKR maturity, we also wanted to understand how OKRs were related to more modern engineering practices. In order to examine this, we identified 7 components of modern engineering (listed below). \added{The components are related to listening to customers, experimentation, rapid iterations, and continuous learning. Safe velocity~\cite{safe} is the concept of moving fast but being able to move safely because of safeguards like flighting, feature gates, telemetry, reporting, alerting, etc. }

\begin{itemize}
\item I do my work with the customer in mind
\item I use experimentation to drive decisions/features
\item I have hypotheses as part of my experiments
\item I work at a safe velocity 
\item I use rapid iterations
\item I mostly commit small changes frequently.
\item I am always learning
\end{itemize}

We then asked individuals how effective they were at each of these practices, using the same Likert scale we used for OKR maturity (a 5 point scale ranging from ``Not effective at all'' to ``Extremely effective'').

\paragraph{OKR Maturity Correlations}
We used linear regression to examine correlations of modern engineering practices with OKR maturity. We found that the practice of using \emph{experimentation to drive decisions/features} and \emph{using rapid iterations} were both positively and significantly correlated with OKR maturity. \added{This is similar to the findings of Ferrazzi which found OKRs complement the agile methodology \cite{ferrazzi}}. The OKR framework stresses having big goals (in fact it says goals should not even be 100\% attainable, but rather aspirational in nature) and regularly checking on progress towards these goals, making adjustments as necessary. We believe having an experimentation mindset aligns well with this big-goal idea. Experiments allow developers to see if an idea will work, and therefore have early proof that it will be successful before shipping broadly. This gives people the confidence to make big, lofty goals, as they know they will have early data to validate or invalidate their idea, allowing them to pivot quickly if something isn't working. The monthly check-in requirements of the OKR framework also align well with making decisions based on experimental data - each month the KRs are evaluated and can be changed if the data suggests it should be. Similarly, quick and on-the-fly changes based on data can only be done through rapid iterations. We believe these two tenets of more modern engineering practices align well in theory with the OKR framework, and that was borne out in our survey data. 

\paragraph{Modern Engineering Score}
Using a similar score mechanism to the OKR framework, we also calculated a \textbf{``Modern Engineering Score''}. We used the 7 items of modern engineering we list above, and scored them using the following equation:

\medskip
Modern Engineering Score:
\begin{equation}
    \sum\limits_{\textit{item}} = 
    \begin{cases}
    1, & \emph{if \emph{item} is \emph{Very Effective} or \emph{Extremely Effective}} \\
    0, & \emph{otherwise}
    \end{cases}
\end{equation}

This score ranges from 0 to 7. We then used linear regression to see how the Modern Engineering Score correlated with each \emph{individual} aspect of using OKRs as well as the demographic and cultural statements explained above. We found that higher scores are correlated with being effective at measuring goals, creativity, focusing on a unified mission, learning and being willing to embrace change. 

\subsection{Unified Mission and Creativity}
Interestingly, having a \emph{unified mission as a team} and having a \emph{team that highly values creativity} were both positively and significantly correlated with \textbf{both} OKR maturity and the Modern Engineering Score. These two tenants of team behavior, therefore, seem to be keys to effective goal setting and solid modern engineering practices.

A unified mission will bring alignment, which is important, as it was previously found that only 7\% of employees ``fully understand their company's business strategies and what's expected of them in order to help achieve common goals'' \cite{kaplanBook}. This alignment might drive people to make better OKRs, as it is clear what they are working toward, allowing individuals to set on-point objectives with relevant Key Results. 

Creativity was the other key component to having both a high Modern Engineering Score and a high OKR Maturity score. Failure is part of the creativity process \cite{PsychologyToday} and so in order to have a creative team there needs to be room for failure. Likely, teams that allow for failure but still succeed must have ways to identify failure early on and pivot. Key components of our Modern Engineering Score are working with an experimentation mindset, always learning and using a hypothesis. These elements allow one to come up with an idea, experiment, and learn from the results. This is essentially the scientific process which also has failure as a fundamental element - it is how we learn. Teams that are strong in this scientific process likely have experienced much failure, know how to handle it and are not phased by it, allowing for creativity to blossom.

These two elements of a team - understanding and working towards a common mission as well as making space for creativity and potential failure - are connected to and possibly drivers of strong OKR culture and goal setting behavior. 

\subsection{Middle Management Matters}
We hypothesized that middle management is likely critical for taking the high level ideas and goals of executives and turning them into actionable projects for individuals. As such, we asked managers about their ability to translate OKRs into goals and metrics for their teams, and also asked both managers and individuals about the cadence of discussion around goals. Somewhat surprisingly more than 50\% of managers reported that they are not very effective at translating OKRs from their management into actionable goals/metrics for their teams. As we saw above, many individuals struggle with not understanding how their work aligns with other teams and leaders; the priorities of the organization; and feelings of frustration around shifting priorities. Often these individuals have little or no facetime with senior leaders and are dependent on their managers for providing information about the goals of the organization and about how their work relates to those goals. As such, managers are critical for the process of taking high level objectives and turning them into actionable key results their team can achieve. 

In addition to a lack of ability, there is also a mismatch in how managers think they are doing and what is really going on with respect to goals. 60\% of individual contributors say that goals are communicated to them monthly, but 65\% of managers say that they communicate goals weekly or multiple times a week. The lack of process around goal setting, and lack of tools, could contribute to this discrepancy. 

\subsection{Remote Teams Report Higher OKR Maturity}
Lastly, we found an interesting correlation when looking at individuals who work away from the head office of this organization. These individuals scored higher on OKR maturity, and in fact a location outside of the main office was positively correlated with higher OKR maturity with a p-value of less than 0.05. We hypothesize this is because teams who do not have regular face time and in person meetings with leadership need to be more clear on what they are doing, how it relates to the org, and have better ability to share back their progress. If a remote individual or team is unclear on what they need to be doing, they could potentially be going off course for weeks before a formal meeting sets them back on course. With people who are co-located, they are more likely to experience serendipitous conversations in the hallway or cafeteria which could illuminate any misunderstandings, and are more likely to be able to sync with people about work over chat (due to time zone similarities). These smaller but more frequent points of contact can help a person gauge their work and adjust as needed. Remote individuals are less likely to have these kinds of interactions. In addition, remote individuals and teams likely need to report back on their progress more proactively to stay top of mind, and as such may have built a stronger goal tracking and communicating muscle.

\section{Top Challenges with using OKRs}
\label{sec:challenges}

\begin{figure}
\includegraphics[width=0.45\textwidth,angle=270,trim={0 1.25cm 0 0},clip]{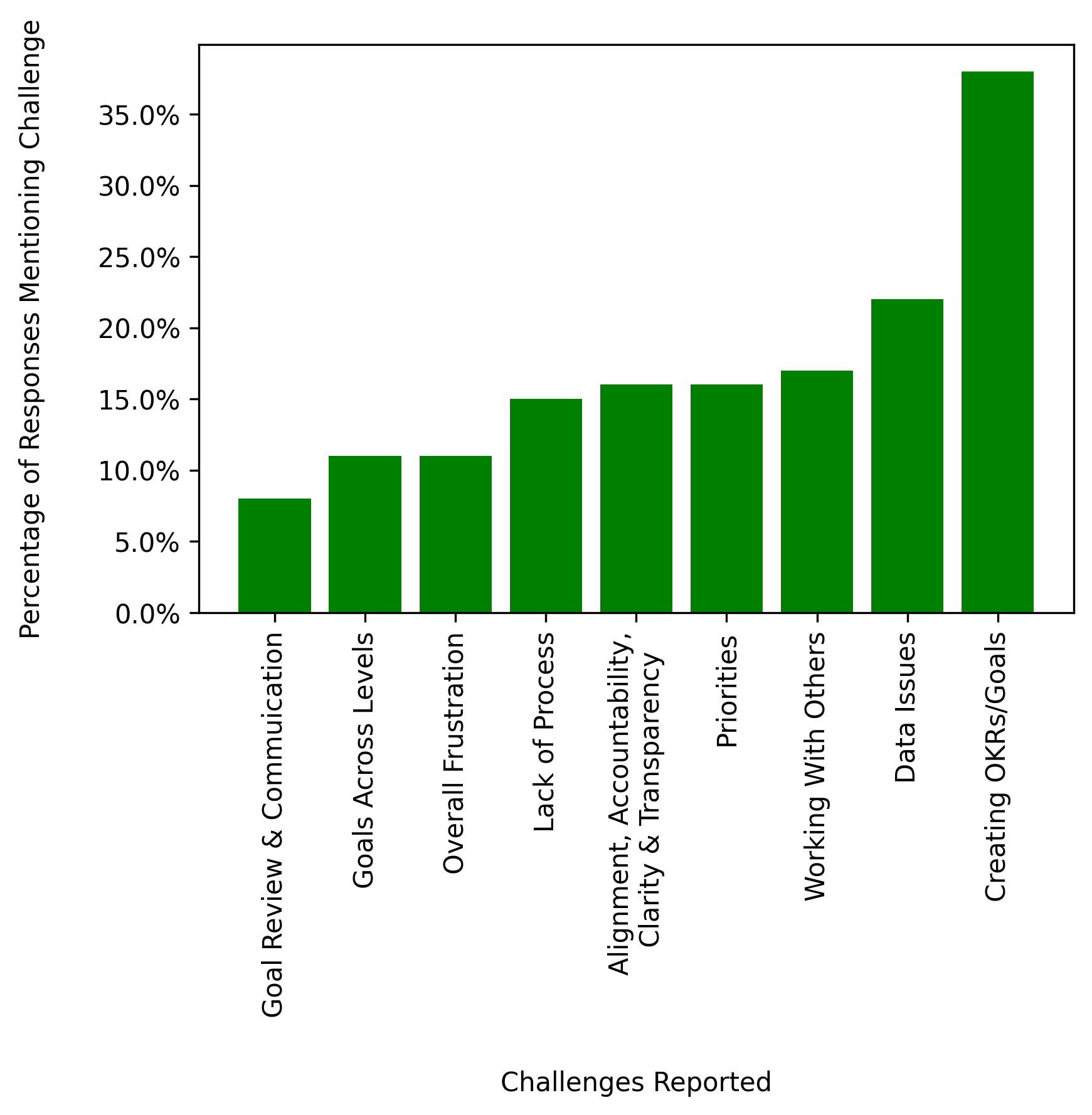}
\caption{Top challenges faced when working with OKRs and goals as reported by survey participants.}
\label{fig:topChallenges}
\end{figure}

We asked respondents what issues they faced when working with OKRs. This was an open text question. We used thematic analysis and open coding to identify the top issues reported. Figure \ref{fig:topChallenges} shows the main challenges found in the text. Please note the total sums to more than 100\% as some people mentioned multiple challenges.

We dug into the top 6 of these challenges and will share highlights of each below.

\subsection{Creating and Setting OKRs}
With almost double the number of occurrences of other challenges, creating and setting OKRs was by the far the most reported challenge people face. People were asked to rate their effectiveness at goal setting, and almost 1 in 5 stated that they do not believe they are effective at all. Indeed, only 45\% of teams think they are effective at measuring goals. Almost one quarter of the challenges mentioned were around creating OKRs in the first place. Here are some verbatims people shared around challenges with creating and setting OKRs:
\newline

\myquote{Select measurable goals that would actually correspond to a real business need and won't promote behavior only to make number nice (ex. contributions would push people to do many very small commits, damaging productivity).}

\myquote{Arbitrary goal setting based on limited data and knowledge. Prioritization based on the arbitrary goals.}

\myquote{Knowing what are appropriate and reasonable goals}

\myquote{Getting started. It is a big culture shift to go from tracking output to outcomes, and our team has significantly improved in this area in the past year.}

As shown in the last verbatim, it seems there is a fair amount of activation energy required to move into an OKR mindset. Historically large software companies may have tracked the work they are doing vs the outcome that work drove. Measuring outcomes (such as new users acquired, sales, etc) can be much harder than just measuring the work one did.

\subsection{Data Issues}

Some kind of issue with data access or data pipe issues were mentioned by more respondents than any other single challenge. Recent changes in how data can be stored and handled (with things such as the \added{General Data Protection Regulation (GDPR) from the EU}, the Executive Order around data in America, etc) means data access is an issue in all engineering teams. Some examples of issues reported are:

\myquote{Getting consistent, trusted data.  You need this to both set and track the KRs.}

\myquote{The many layers of tech required to gather data and build actionable reports. You need to know a lot about a lot of different tech stacks.}

\myquote{Tracking goals is a bit harder because our data systems are not always set up to be conducive for this -- it can be frustrating to figure out how to get data to track a particular goal, and all of the privacy/security hoops that we have to jump through are infuriating. Reporting OKRs isn't too bad, but sometimes the challenges of tracking OKRs can also make the reporting more complicated.}

As stated above, data is key to setting and tracking OKRs, but in this large organization getting access to data was a serious pain point. Companies need to comply with a variety of data requirements, but they can still attempt to set up their data pipeline in a way that makes it easier for individual contributors to meaningfully measure progress. Using a smaller set of tools, thinking about the data needed for a feature before feature creation, and adequate training on data access can all help alleviate the challenges of working with data.

\subsection{Working With Others}

One of the superpowers of OKRs is the ability to ``Align and Connect for Teamwork''. In theory, when all OKRs are shared broadly and transparently, teams can see dependencies or accidental duplication of work early and correct. In this large organization with many teams, working with others continues to be a challenge. This team builds a large piece of software that has existed for many years, with some teams being new, some old, and some acquired, making for many different ways of working. Ideally aligning on a single OKR process and tool will help these teams work together, but at the time of the study working across teams still proved difficult. Some specific challenges people reported are below: 

\myquote{It is difficult to get alignment at my manager and skip manager's level on what are the actual outcomes we are trying to drive.  This makes it more difficult to work on cross-team projects where the defined OKRs are slightly different and each team has tunnel vision about their specifically defined metrics.}

\myquote{Working with partner teams that have old-school processes that involve a lot of paperwork (they want formal specs, lengthy OKR definitions, reviews and scorecards every week, etc.). There is SO much overhead with even just engaging at the surface-level with those teams that it completely deflates the morale of everyone on our team with wanting to engage with them (but of course we keep on doing it because it's what's best for the product).}

\myquote{Cross-functional efforts are difficult to quantify and measure especially those which have dependencies on other teams work. The goal tracking and measuring is broken because of: a) No standard way of tracking it. (We use Onenote, daily/weekly scrum, online tools and eventually Outlook) and b) Because of no standard way of tracking it, it is difficult to align your day to day work to a certain goal.}

\subsection{Priorities} 
The first superpower of OKRs is ``Focus and Commit to Priorities''~\cite{Doerr01}. When the OKR system is used properly, everyone can see each other's commitments and work, and individual contributors can see their work connect up to top level goals of the organization. \added{In our study of this organization that was early in the process of using OKRs, there were considerable struggles with priorities}. Individuals felt like leadership selects a priority and then has the ability to change it at will. Aligning work to leadership priorities, then, becomes a futile exercise. In addition, they found it hard to align on a priority as a team. If individuals publicly state what they are working on, this could lead to them being held accountable for that, and people did not necessarily want that. Without clear commitment from the leadership of what the priorities are and that they will be stuck to, sticking to priorities at a lower level felt futile.

\myquote{Unifying the team behind the goal. Individuals don't necessarily want to be held accountable. This is especially difficult when [the leadership team] is not as actively engaged in the OKR process at the individual team level.}

\myquote{Leadership getting into agreement as to what the priorities should be (each individual org seems to have their own agenda)}

\subsection{Alignment, Accountability, Clarity and Transparency}

The 2nd most common individual challenge people reported as lacking clarity and alignment within their organization. As mentioned above, this partly comes from a lack of process around ORKs, but people report not just a lack of understanding around the process of OKRs but about the goals of the organization themselves. In Measure What Matters John Doerr states ``by clearing the line of sight to everyone's objectives, OKRs expose redundant efforts and save time and money''~\cite{Doerr01}. Ideally, OKRs are shared broadly and transparently within an organization, so everyone can see each person's commitment work - alignment is actually supposed to be a super power of the OKR system! Within this organization, a lack of a central tool or process meant people did not know what work was being done around the organization without doing leg work to learn. In fact, on the survey we asked an open text question of ``What one thing would you change or try that you believe would help with goal setting and achievement in [organization]?'' and a top answer was clear, focused OKR goals and priorities from the organization leadership. Here are some verbatims that show the current struggles people face with alignment and clarity:

\myquote{Knowing how what I do daily ties in to larger goals. Knowing what might be tracked, and how it is tracked, so that I can make sure what I do gets reported. Turning what I do daily into something that can and is tracked.}

\myquote{Lack of clarity about how my goals align with broader goals and whether I should be shifting to match management's goals more.}

\myquote{My experience has been that when there is misalignment of goals among two organizations, that can lead to less productive time use on the team that has made the shared goal a higher priority. From the IC level this is frustrating, though I get that aligning everyone is a huge challenge.}

\subsection{Lack of Process and Tools}

A key component of the OKR framework is that the OKRs are open and transparent - anyone in the team should be able to see other people's OKRs. Our survey found that 76\% of people don't know how to find other teams' goals when they need them, and 59\% of people said they want or need to see other teams' OKRs. This was largely because of a lack of process and tools. Like many large software companies, this team had been around for many years, and different parts of the team have different styles of work. This flowed into their OKR style as well - each team took the concept and implemented it as they wanted, without a central process or tool. Because of this, other teams did not know how to view or interact with OKRs outside of their team, and sometimes even within teams. \added{In fact, it was reported that 12 different tools were used for OKR tracking in this one large organization}. A lack of process from the top down was also found. This meant individuals did not always understand why they were measuring and tracking goals, or how it related or influenced the overall direction of the organization (something OKRs are meant to address). Some verbatims that demonstrate this lack of process and tools are:

\myquote{From my perspective, every team/org does this differently. It would be more efficient if we worked together to really have a refined strategy on OKRs (including tooling)} 

\myquote{The whole goal setting process seems very nebulous and disjointed... What are the key results we're aiming for? How will we measure success? But my biggest struggle is that there doesn't appear to be a consistent way to set goals for customer scenarios that require work across multiple orgs. There's too much focus on optimizing locally instead of globally.}

\section{Recommendations}
\label{sec:rec}

Based on the challenges reported, the correlations with high OKR maturity, and the open-text suggestions from the survey, we have come up with a set of recommendations for improving the goal setting and OKR process in software organizations:
\begin{itemize}
    \item Invest in the data pipeline
    \item Increase transparency
    \item Improve communication
    \item Promote learning communities 
    \item Guide the process
\end{itemize}

\subsection{Invest in the Data Pipeline}
A key part of the OKR framework is setting and measuring goals. This is effectively impossible without access to data. Even when data exists, if there are barriers to accessing that data people will often not use it. We saw in the verbatims that sometimes accessing data was so challenging a person would do a mental-math trade-off on whether it was even worth it to instrument the feature with data in the first place. Here is what one person had to say about this:

\myquote{Since there's this huge cost to measuring actual customer signals, I've tried to get better at minimizing when we do all this. A lot of the guidance we've heard was to measure+experiment on every small thing, but with so much overhead, it just doesn't make sense. The ballpark measure I tend to use is, if we're going to spend ~1 month on creating/managing these telemetry signals over the course of feature development, it has to pay that back somehow.}

Investing in an easy to use, compliant data pipeline will allow for more data-driven OKRs which makes the whole framework more effective.

\subsection{Increase Transparency}
A lack of understanding the goals and work of other teams was seen in a variety of the top challenges. This caused issues with alignment, working with other teams, understanding how one's work relates to the broader org and flat-out frustration and lack of belief in the process. When OKRs are shared transparently teams understand what others are working on and proactively align their work properly. They also can see what matters to their leadership and position their work respectively. For this organization having a single place to store OKRs would greatly improve transparency. In fact, since this study was conducted, the organization has embarked on a mission to move all teams into a single OKR management tool. It is a future area of research to understand how moving to this tool impacted the success of the OKR process in the organization.

\subsection{Improve Communication}
As we saw, managers did not feel very effective at taking high level ideas and translating them into work their teams could do. We also saw a large discrepancy between how much a manager thinks they are communicating about their goals and how much their team actually receives. Prior work has shown that facilitating external communication, driving alignment and guiding a team are key elements that make a great software engineering manager \cite{greatmanager}. These same skills need to be present for a manager to effectively lead goal setting and goal alignment. Managers need to especially focus on driving clarity about how the work of individuals and their team relates to both other teams and the overall goals of the organization.

\subsection{Promote Learning Communities}
It was found that some teams do indeed have high OKR maturity. We also saw in the verbatims that it is challenging for more modern teams to work with teams that still use older methodologies. Since cross-team issues are so prevalent, it would be wise to form learning communities that bring people together across different teams. Ideally, teams with a strong OKR process could be paired with teams who are still new to the process. They can observe how the team works, share best practices, and help guide the new team. This would have the dual benefit of helping less mature teams improve and helping more mature teams be able to partner with other mature teams, as they would be helping a team mature. Teams could also appoint an \emph{``OKR Champ''} to oversee the process, and more experienced champs could mentor new champs. These learning communities could share best practices which would also help drive alignment throughout the organization as more teams would be following a similar OKR procedure.

\subsection{Guide the Process}
This organization has existed for decades, and as mentioned above, consists of a mix of teams from around the world - some long standing, some new, some acquired. Each team works in their own way - planning, tracking and sharing out information in whatever way works best for them. When some teams started using OKRs it was not due to a decision the organization made - rather, some individuals had heard about OKRs or used them at a prior job and wanted to bring the goodness of that system to their team. Pockets of OKRs started to appear. At the high leadership level, some leaders decided it would be good, and started to use it as well. Up until this point, there had been no official role out of this framework. Teams chose to use it, or not, and chose how to use it - what tool, at what cadence, etc. This made goal setting and tracking across teams extremely difficult. OKRs are actually defined by Niven and Lamorte as a ``a critical thinking framework and ongoing discipline that seeks to ensure employees work together, focusing their efforts to make measurable contributions that drive the company forward.''~\cite{lamorte}. It is meant to bring people together and to be used as an ongoing discipline. This requires people to work together and in a similar way in order to be productive. For this framework to be useful in a software organization, it must be used consistently across teams and should be rolled out in a controlled manner, with experts leading the rollout and implementing some of the strategies above (setting up learning communities, ensuring data is available, etc).

\section{Limitations}

\added{In this section, we discuss limitations of this work.

This research relied on self-reported data, both through interviews and surveys, which introduces a limitation due to potential biases like social desirability and recall bias~\cite{podsakoff2003common}. 
Despite these biases, self-reported data is often deemed valuable and pragmatically viable in organizational research, especially when objective measures are elusive or impractical~\cite{conway2010reviewers}. 
The validity of such data can be substantiated when consistency is observed across diverse respondents and when it correlates with related variables~\cite{donaldson2002understanding}, as we've demonstrated in this study.
Thus, while recognizing its potential shortcomings, we believe that the use of self-reported data is the most viable approach to gain insights into both the implementation and maturity of OKRs.

The questions in the survey we deployed were original and not previously validated due to the absence of any pre-existing, validated instrument for exploring OKRs.
While this approach enabled us to tailor our investigation to our specific research context, it may impact the generalizability of our findings. 
We share these questions in supplemental materials~\cite{supplemental-materials} as a starting point for future research, inviting subsequent studies to validate and refine them, thereby contributing to the development of robust research tools in the OKR domain.

The actual implementation of OKRs for an organization will be affected by many contextual factors. 
Our findings represent one set of such factors but may not generalize to companies of differing size (e.g., 100,000 vs. 20 developers) or domain, etc., due to the varying challenges, organizational cultures, and priorities.
Consequently, while our insights provide valuable perspectives on OKR implementation within the context of our study, one should consider contextual similarities and differences when extrapolating our findings to other organizations. }

Finally, this paper presents a case study of just one company.
Some believe that that empirical research within 
one company provides little value for the academic community and does not contribute to 
scientific development. Historical evidence shows that this is not the case. 
Flyvbjerg provides several examples of individual cases that contributed to discovery in physics, economics, and social science~\cite{flyvbjerg2006five}.
Beveridge observed for social sciences: “More discoveries 
have arisen from intense observation than from statistics applied 
to large groups” (as quoted in Kuper~\cite{kuper1985}, page 95). 
Please note that this should not be interpreted as a criticism of 
research that focuses on large samples or entire populations. For 
the development of an empirical body of knowledge as championed by Basili~\cite{basili1999building}, both types of research are essential. The 
work presented in this paper comes from a study of one company with tens of thousands of developers, as well as many different development methodologies, domains, and processes.


\section{Conclusion}
As we saw, in their book ``Objectives and Key Results'', Niven and Lamorte call out that objectives and key results are not just a goal setting framework, but rather a critical thinking framework as well as an ongoing discipline. Doerr suggests that using this framework and abiding by the discipline will provide teams with four superpowers: 1) focus and commit the priorities; 2) align and connect for teamwork; 3) track for accountability; and 4) stretch for amazing. What we have found in this research is that the promise of these superpowers is possible, however using the framework in a software engineering setting is challenging and requires oversight.

Many engineering teams already use software for tracking their work (such as Azure Dev Ops) and have other work specific tools. Engineers are also very good at finding, or even building, creative solutions to problems. As such, implementing an OKR framework with a large group of software engineers without careful rollout and tooling will let a thousand flowers bloom - teams will find their own, creative, unique way to use this process. In these cases, problems such as a lack of clarity and alignment, difficulty setting and sharing out goals, challenges working with other teams, and shifting priorities are bound to come up.

Rather, in order to successfully use OKRs in the software industry, managers need to be properly trained on the framework; learnings need to be shared and consistent; data needs to be accessible; and goals at all levels need to be transparent and clearly communicated. When teams have a high OKR maturity, they are able to set and measure goals well, as well as reflect on progress towards goals and change course as needed. This maturity is correlated with other desirable skills for software engineers, including creativity, team commitment to to organizational priorities, enjoying the work they are doing and being overall happy at the company (including a decreased desire to leave). In addition, having a unified mission as a team and valuing creativity are correlated with both high OKR maturity, and more modern development practices, something many existing software engineering teams are trying to pivot to. It is clear that objectives and key results can be beneficial for software engineering teams when used consistently across teams and organizations.

%
\begin{acks}
We thank all participants in the interviews and the survey for their insightful comments on working with OKRs and goals.
We would also like to thank \added{Paul Bennett, Nicole Forsgren, Cory Hilke, Ales Holecek, Ron Pessner, Josh Pollock, David Speirs, Jaime Teevan, and Igor Zaika as well as the entire SAINTES team} for their constructive and insightful feedback throughout all stages of this project.
The ethics for this study were reviewed and approved by the Microsoft Research Institutional Review Board (MSRIRB), which is an IRB federally registered with the United States Department of Health \& Human Services. 
\end{acks}

%
\bibliographystyle{ACM-Reference-Format}
\bibliography{ref}

%

\end{document}

%% file: utils.tex
\usepackage{pgfplots}
\pgfplotsset{width=10cm,compat=1.9}

\usepackage{stackengine}
\usepackage{verbatim}
\usepackage{xspace}

\usepackage{xcolor}
\definecolor{darkgreen}{rgb}{0.05,0.5,0.05}

\newcommand{\added}[1]{{#1}\xspace}

\renewenvironment{quote}{%
   \list{}{%
     \leftmargin\parindent
     \rightmargin0cm
   }
   \item\relax
}
{\endlist}

\usepackage{fontawesome5}
\newcommand{\myquote}[1]{\begin{quote}{\small\faComment}~\small\emph{#1}\end{quote}}

